\documentclass[12pt,a4paper]{article}
\usepackage{epsfig}
\begin{document}
\title{Neutral top-pion and lepton flavor violating processes}
\author{Chongxing Yue$^{a}$, Dongqi Yu$^{a}$, Lanjun Liu$^{b}$\\
{\small a: Department of Physics, Liaoning  Normal University,
Dalian 116029. P.R.China}
\thanks{E-mail:cxyue@lnnu.edu.cn}
\\ {\small b: College of Physics and Information Engineering,}\\
\small{Henan Normal University, Xinxiang  453002. P.R.China}
 }
\date{\today}
\maketitle

\begin{abstract}
\hspace{5mm}In the context of topcolor-assisted techicolor(TC2)
models, we study the contributions of the neutral top-pion
$\pi^{0}_{t}$ to the lepton flavor violating(LFV) processes
$l_{i}\rightarrow l_{j}\gamma$ and $l_{i}\rightarrow
l_{j}l_{k}l_{l}$. We find that the present experimental bound on
$\mu\rightarrow e\gamma$ gives severe constraints on the free
parameters of $TC2$ models. Taking into account these constraints,
we consider the processes $l_{i}\rightarrow l_{j}l_{k}l_{l}$
generated by top-pion exchange at the tree-level and the one loop
level, and obtain $Br(\mu\rightarrow 3e)\simeq 2.87\times
10^{-14}$, $1.1\times 10^{-15}\leq Br(\tau\rightarrow 3e)\simeq
Br(\tau\rightarrow 2e\mu)\leq 4.4 \times 10^{-15} $, $3.1\times
10^{-15} \leq Br(\tau\rightarrow 2\mu e)\simeq Br(\tau\rightarrow
3\mu)\leq 1.5 \times 10^{-14}$ in most of the parameter space.
\end {abstract}

\vspace{2.0cm} \noindent
 {\bf PACS number(s)}:12.60.Cn, 13.35.Dx, 14.80.Mz, 14.80.Cp

\newpage

\vspace{.5cm} \noindent{\bf 1. Introduction}

The standard model $(SM)$ accommodates fermion and weak gauge
boson masses by including a fundamental scalar
$Higgs\hspace{0.2mm}H.$ However, the $SM$ can not explain the
dynamics responsible for the generation of mass. Furthermore, the
scalar sector suffers from the problems of triviality and
unnaturalness. Thus, the $SM$ can only be an effective field
theory below some high-energy scale. New physics should exist at
energy scales around $TeV$.

The cause of electroweak symmetry breaking $(EWSB)$ and the origin
of fermion masses are important problems of current particle
physics. Given the large value of the top quark mass and the
sizable splitting between the masses of the top and bottom quarks,
it is natural to wonder whether $m_{t}$ has a different origin
from the masses of other quarks and leptons. There may be a common
origin for $EWSB$ and top quark mass generation. Much theoretical
work has been carried out in connection to the top quark and
$EWSB$. Topcolor-assisted technicolor $(TC2)$ models [1],
flavor-universal $TC2$ models [2], top see-saw models [3], and top
flavor see-saw models [4] are four of such examples. The common
feature of these kinds of models is that topcolor interactions are
assumed to be chiral critically strong at the scale about 1 $TeV$,
and it is coupled preferentially to the third generation. In TC2
models, $EWSB$ is mainly generated by $TC$ interactions or other
strong interactions. The topcolor interactions also make small
contributions  to $EWSB$ and give rise to the main part of the top
quark mass. Then, the presence of the physical top-pions in the
low-energy spectrum is an inevitable feature of these kinds of
models. Thus, studying the possible signatures of the top-pions at
present and future high- or low- energy colliders can help the
collider experiments to search for top-pions, test topcolor
scenario and further to probe $EWSB$ mechanism.

It is well known that the individual lepton numbers $L_{e},
L_{\mu}$, and $L_{\tau}$ are automatically conserved and the
tree-level lepton flavor violating $(LFV)$ processes are absent in
$SM$, due to unitary of the leptonic analog of $CKM$ mixing matrix
and the masslessness of the three neutrinos. However, the solar
neutrino experiments [5] and the atmospheric neutrino experiments
[6] confirmed by reactor and accelerator experiments [7] provide
very strong evidence for mixing and oscillation of the flavor
neutrinos, which presently provide the only direct observation of
physics that can not be accommodated within the $SM$ and imply
that the separated lepton number are not conserved. Thus, the $SM$
requires some modification to account for the pattern of neutrino
mixing, in which the $LFV$ processes like $l_{i}\rightarrow
\l_{j}\gamma$ and $l_{i}\rightarrow l_{j}l_{k}l_{l}$ are allowed.
The observation of these $LFV$ processes would be a clear
signature of new physics beyond the $SM$. The fact and the
improvement of their experimental measurements force one to make
more elaborate theoretical calculation in the framework of some
specific models beyond the $SM$ and see whether the $LFV$ effects
can be tested in the future experiments. For instance, these $LFV$
processes have been widely studied in a model independent way in
Ref.[8], in the $SM$ with extended right-handed and left-handed
neutrino sectors [9], in supersymmetric models [10], in the
general two $Higgs$ doublet model($2HDM$) type III [11], in the
$Zee$ model [12], and in the topcolor models [13].

The aim of this paper is to study the contributions of the neutral
top-pion $\pi^{0}_{t}$ predicted by $TC2$ models to the $LFV$
processes $l_{i}\rightarrow l_{j}\gamma$ and $l_{i}\rightarrow
l_{j}l_{k}l_{l}$ and see whether $\pi^{0}_{t}$ can give
significant effects on these processes. The paper is organized as
follows: in section 2 we give the flavor-diagonal(FD) and
flavor-changing(FC) couplings of $\pi^{0}_{t}$ to the three family
leptons and calculate its contributions to the $LFV$ process
$l_{i}\rightarrow l_{j}\gamma$. Using the experimental upper limit
of the $LFV$ process $\mu\rightarrow e\gamma$, we give the
constraints on the flavor mixing factors. The tree-level and one
loop-level contributions of $\pi^{0}_{t}$ to the branching ratios
$Br(l_{i}\rightarrow l_{j}l_{k}l_{l})$ are calculated in section
3. The conclusions are given in section 4.

\vspace{.5cm} \noindent{\bf 2. The flavor mixing factors of
$\pi^{0}_{t}$ and the $LFV$ processes $l_{i}\rightarrow
l_{j}\gamma$}

2.1 The couplings of the neutral top-pion $\pi^{0}_{t}$ to
leptons.

For TC2 models [1], $TC$ interactions play a main role in breaking
the electroweak symmetry. Topcolor interactions make small
contributions to $EWSB$, and give rise to the main part of the top
quark mass, $(1-\varepsilon)m_{t}$, with the parameter
$\varepsilon\ll1$. Thus, there is the following relation:
\begin{equation}
\nu^{2}_{\pi}+F^{2}_{t}=\nu^{2}_{W},
\end{equation}
where $\nu_{\pi}$ represents the contributions of $TC$
interactions to $EWSB$, $\nu_{W}=\nu/\sqrt{2}\simeq 174GeV$. Here
$F_{t}\simeq 50GeV$ is the physical top-pion decay constant, which
can be estimated from the Pagels-Stokar formula. This means that
the masses of the gauge bosons $W$ and $Z$ are given by absorbing
the linear combination of the top-pions and technipions. The
orthogonal combination of the top-pions and technipions remains
unabsorbed and physical [14]. However, the absorbed Goldstone
linear combination is mostly the technipions while the physical
linear combination is mostly the top-pions, which are usually
called physical top-pions ($\pi^{\pm}_{t},\pi^{0}_{t}$). The
existence of the physical top-pions in the low-energy spectrum can
be seen as characteristic of topcolor scenario, regardless of the
dynamics responsible for $EWSB$ and other quark masses.

The $FD$ couplings of the neutral top-pion $\pi^{0}_{t}$ to
leptons can be written as:
\begin{equation}
\frac{m_{l}}{\nu}\overline{l}\gamma^{5}l\pi^{0}_{t}
\end{equation}
where $l=\tau$, $\mu$ or $e$. For $TC2$ models, the underlying
interactions, topcolor interactions, are non-universal and
therefore do not posses $GIM$ mechanism. The non-universal gauge
interactions result in the new $FC$ coupling vertices when one
writes the interactions in the mass eigen-basis. Thus, the
top-pions can induce the new $FC$ scalar coupling vertices [15].
The $FC$ couplings of $\pi^{0}_{t}$ to leptons can be written as:
\begin{equation}
\frac{m_{\tau}}{\nu}k_{\tau
i}\overline{\tau}\gamma^{5}l_{i}\pi^{0}_{t},
\end{equation}
where $l_{i}(i=1,2)$ is the first(second) lepton e($\mu$),
$k_{\tau i}$ is the flavor mixing factor, which is the free
parameter.

Using the $FD$ and $FC$ couplings of the neutral top-pion
$\pi^{0}_{t}$ to fermions, we have studied the $FC$ process
$\mu^{+}\mu^{-}\rightarrow\overline{t}c$ mediated by
$\pi^{0}_{t}$ exchange [16]. We find that $\pi^{0}_{t}$ can
generate several hundred $\overline{t}c$ events and the signals of
$\pi^{0}_{t}$ might be detected via the process
$\mu^{+}\mu^{-}\rightarrow\overline{t}c$ in the first muon
collider. In the next subsection, we will study the contributions
of $\pi^{0}_{t}$ to the $LFV$ processes $l_{i}\rightarrow
l_{j}\gamma$ and see whether the experimental upper limits of
these processes can give significant constraints on the flavor
mixing factor $k_{\tau i}$.

2.2 The $LFV$ processes $l_{i}\rightarrow l_{j}\gamma$

The observation of neutrino oscillations [5,6] implies that
individual lepton numbers $L_{e,\mu,\tau}$ are violated,
suggesting the appearance of the $LFV$ processes, such as
$l_{i}\rightarrow l_{j}\gamma$ and $l_{i}\rightarrow
l_{i}l_{j}l_{l}$. The branching ratios of these processes are
extremely small in the $SM$ with right-handed neutrinos. For
example, reference [17] has showed $Br(\mu\rightarrow
e\gamma)<10^{-47}$. Such small branching ratios are unobservable.

\begin{figure}[htb]
\vspace{-9cm}
\begin{center}
\epsfig{file=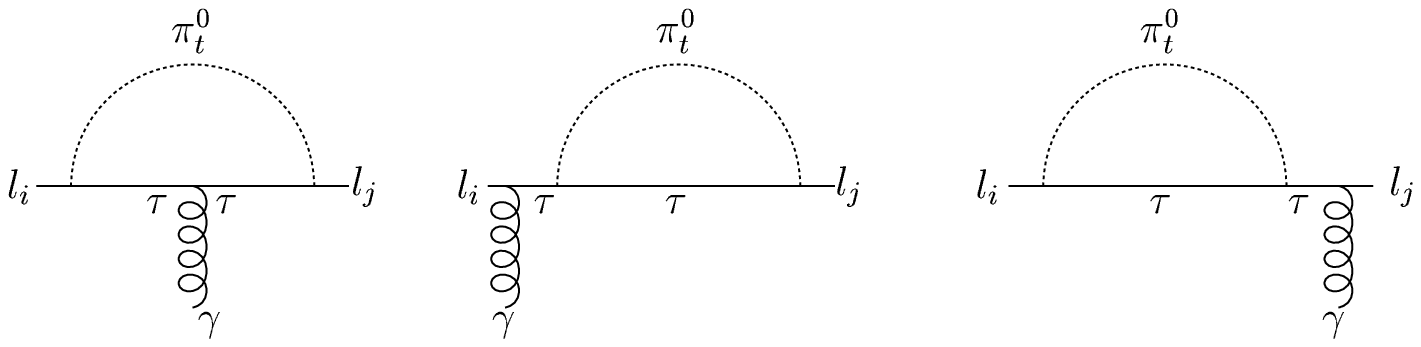,width=550pt,height=800pt} \vspace{-16.5cm}
\hspace{1cm} \caption{Feynman diagrams contribute to the $LFV$
processes $l_{i}\rightarrow l_{j}\gamma$ due to the neutral
\hspace*{1.8cm} top-pion $\pi^{0}_{t}$ exchange in $TC2$ models.}
\label{ee}
\end{center}
\end{figure}

The present experimental upper limits are [18]:
\begin{eqnarray}
Br(\tau\rightarrow\mu\gamma)&<&1.1\times10^{-6},\nonumber\\Br(\tau\rightarrow
e \gamma)&<&2.7\times10^{-6},\nonumber\\Br(\mu\rightarrow e
\gamma)&<&1.1\times10^{-11}.
\end{eqnarray}
These bounds are expected to be improved by a few orders of
magnitude in the future. For example, in an experiment under
preparation at $PSI$ [19], it is planned to reach a sensitivity
\begin{equation}
Br(\mu\rightarrow e \gamma)<1\times10^{-14}.
\end{equation}
Thus, these processes are ideal tools to search for new physics.
The observation of any rate for one of these processes would be a
signal of new physics.

The relevant Feynman diagrams for the contributions of the neutral
top-pion $\pi^{0}_{t}$ to the $LFV$ processes $l_{i}\rightarrow
l_{j}\gamma$ are shown in Fig.1. The internal fermion lines may be
$\tau,\mu,$ or $e$. However, the internal fermion propagator
provides a term proportional to $m^{2}_{f}$ in the numerator,
which is not cancelled by the $m^{2}_{f}$ in the denominator since
the heavy $\pi^{0}_{t}$ mass dominates the denominator. Thus, we
only take the internal fermion line as the $\tau$ fermion line.

Using Eq.(2), Eq.(3) and other relevant Feynman rules, the decay
widths of the $LFV$ processes $l_{i}\rightarrow l_{j}\gamma$ can
be written as:
\begin{equation}
\Gamma(\tau\rightarrow m\gamma)=\frac{m_{\tau}^{5}k^{2}_{\tau
m}\alpha_{e}}{2048\nu^{4}\pi^{4}}[F^{2}_{1}-\frac{1}{2}
m^{2}_{\tau}(F^{2}_{2}+F_{2}F_{3})-m_{\tau}F_{1}F_{2}],
\end{equation}
where $m =\mu$, or $e$, $F_{i}$ are
\begin{eqnarray}
F_{1}&=&B_{0}+m^{2}_{\pi_{t}}C_{0}-2C_{24}+m_{\tau}^{2}(C_{11}-C_{12})-B^{*}_{0}-B^{'}_{1},
\\F_{2}&=&2m_{\tau}(-C_{21}-C_{22}+2C_{23}),\\F_{3}&=&2m_{\tau}(C_{22}-C_{23}).
\end{eqnarray}
\begin{equation}
\Gamma(\mu\rightarrow
e\gamma)=\frac{m_{\tau}^{4}m_{\mu}k^{2}_{\tau \mu}k^{2}_{\tau e
}\alpha_{e}}{2048\nu^{4}\pi^{4}}[F'^{2}_{1}-m_{\mu}F'_{1}F'_{2}-
\frac{1}{2}m^{2}_{\mu}(F'^{2}_{2}+F'_{2}F'_{3})]
\end{equation}
with
\begin{eqnarray}
F^{'}_{1}&=&m^{2}_{\mu}(C_{11}-C_{12})+m_{\tau}(m_{\mu}-m_{\tau})C_{0}-2C_{24}+
B_{0}\nonumber\\
&&\hspace{0.5cm}+m^{2}_{\pi_{t}}C_{0}-\frac{m_{\tau}}{m_{\mu}}B^{*}_{0}+
\frac{m_{\tau}-m_{\mu}}{m_{\mu}}B^{'}_{0}-B^{'}_{1},\\
F^{'}_{2}&=&2[(m_{\tau}-m_{\mu})(C_{11}-C_{12})-m_{\mu}(C_{21}+C_{22}-C_{23})],\\
F^{'}_{3}&=&2[m_{\tau}C_{12}+m_{\mu}(C_{22}-C_{23})].
\end{eqnarray}
The standard Feynman integrals $B_{n}$, $C_{0}$, and $C_{ij}$ can
be written as:

\begin{equation}
C_{ij}=C_{ij}(-p_{l},p_{\gamma},m_{\pi_{t}},m_{\tau},m_{\tau}),
\hspace{0.5cm}C_{0}=C_{0}(-p_{l},p_{\gamma},m_{\pi_{t}},m_{\tau},m_{\tau}),
\end{equation}
\begin{equation}
B_{0}=B_{0}(p_{\gamma},m_{\tau},m_{\tau}),
\hspace{0.5cm}B^{*}_{0}=B_{0}(-p_{m},m_{\pi_{t}},m_{\tau}),\hspace{0.5cm}
B^{'}_{1}=B^{'}_{1}(-p_{l},m_{\pi_{t}},m_{\tau}),
\end{equation}
where $ m_{\pi_{t}}$is the mass of the top-pion, the variable
$P_{m}$($m=\mu$ or $e$) is the momentum of the find state lepton,
$P_{l}$ is the momentum of the initial state lepton, and $l=\tau$
or $\mu$, which corresponds the lepton $\tau$ decay
$\tau\rightarrow e(\mu ) \gamma$ and the lepton $\mu$ decay
$\mu\rightarrow e \gamma$, respectively. In above equations, we
have assumed that the masses of the final state lepton equal to
zero.

\begin{figure}[htb]
\vspace{-0.5cm}
\begin{center}
\epsfig{file=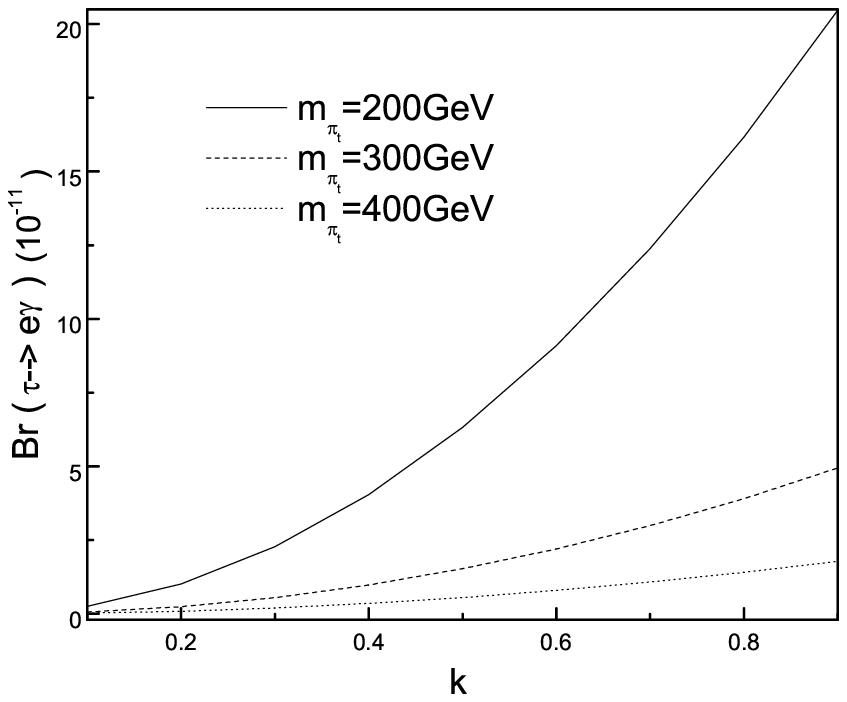,width=350pt,height=300pt} \vspace{-1.0cm}
\hspace{5mm} \caption{The branching ratio $Br(\tau\rightarrow
e\gamma)$ as a function of the flavor mixing factor $k$ for
\hspace*{1.8cm}three values of the top-pion mass $m_{\pi_{t}}$.}
\label{ee}
\end{center}
\end{figure}

 For $TC2$ models, the topcolor interactions only contact with the
third generation. The new particles, such as gauge boson $Z^{'}$
and top-pions $\pi^{0,\pm}_{t}$, treat the fermions in the third
generation differently from those in the first and second
generation and treat the fermions in the first generation same as
those in the second generation. So, in our calculation, we will
assume that the mixing factor $k_{\tau\mu}$ is equal to the mixing
factor $k_{\tau e}$. In this case, we have
$\Gamma(\tau\rightarrow\mu\gamma)\simeq\Gamma(\tau\rightarrow
e\gamma)$ for $m_{\mu}\simeq0,m_{e}\simeq0$. The corresponding
branching ratios $Br(l_{i}\rightarrow l_{j}\gamma)$ can be written
as:
\begin{eqnarray}
Br(\tau\rightarrow \mu\gamma)&\simeq&Br(\tau\rightarrow
e\gamma)=Br^{exp}(\tau\rightarrow e \nu_{e}\overline{\nu}_{\tau})
\frac{\Gamma(\tau\rightarrow e\gamma)}{\Gamma(\tau\rightarrow
e\nu_{e}\overline{\nu}_{\tau})},\\
Br(\mu\rightarrow e\gamma)&=&\frac{\Gamma(\mu\rightarrow
e\gamma)}{\Gamma(\mu\rightarrow e\nu_{e}\overline{\nu}_{\mu})}
\end{eqnarray}
with
\begin{eqnarray}
\Gamma(\tau\rightarrow
e\nu_{e}\overline{\nu}_{\tau})&=&\frac{m^{5}_{\tau}G^{2}_{F}}{192\pi^{3}},\hspace{1.5cm}
\Gamma(\mu\rightarrow
e\nu_{e}\overline{\nu}_{\tau})=\frac{m^{5}_{\mu}G^{2}_{F}}{192\pi^{3}}.
\end{eqnarray}
Where the Fermi coupling constant
$G_{F}=1.16637\times10^{-5}GeV^{-2}$ and the precision measured
branching ratio $Br^{exp}(\tau\rightarrow e
\nu_{e}\overline{\nu}_{\tau})=(17.83\pm 0.06)\%$ [18].

\begin{figure}[htb]
\vspace{0.5cm}
\begin{center}
\epsfig{file=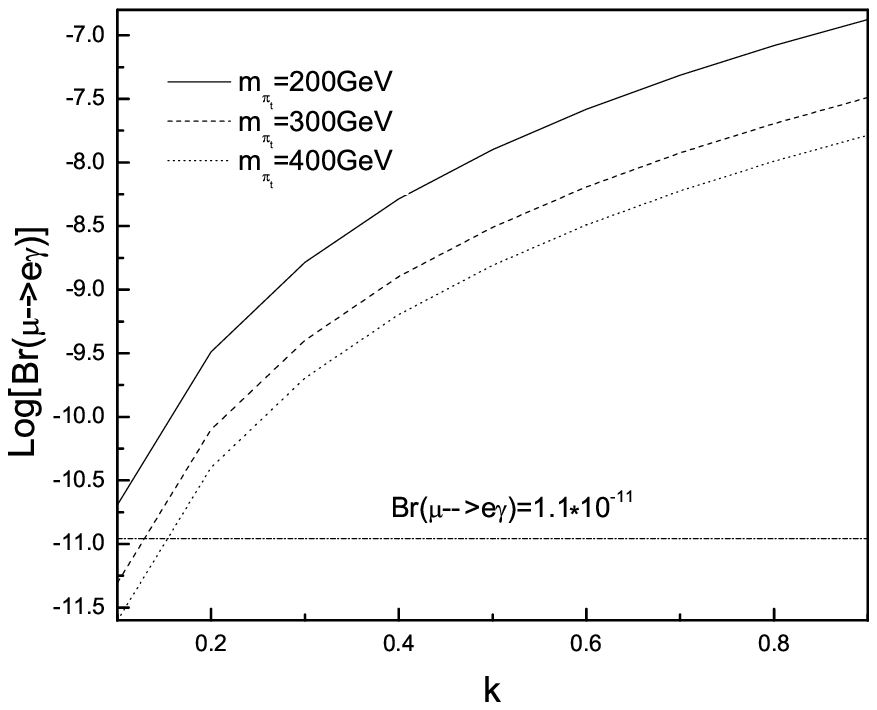,width=350pt,height=300pt} \vspace{-1.0cm}
\hspace{5mm} \caption{The branching ratio $Br(\mu\rightarrow
e\gamma)$ as a function of the flavor mixing factor $ k$
\hspace*{1.8cm} for three values of the top-pion mass
$m_{\pi_{t}}$.} \label{ee}
\end{center}
\end{figure}

 To obtain numerical results, we take the $SM$ parameters as
$\alpha_{e}(m_Z)=\frac{1}{128.8}$, $m_{\tau}=1.777GeV$,
$m_{\mu}=0.105GeV$ [18]. The limits on the mass $m_{\pi_{t}} $ of
the top-pion may be obtained via studying its effects on various
experimental observables [20]. It has been shown that
$m_{\pi_{t}}$ is allowed to be in the range of a few hundred $GeV$
depending on the models. As numerical estimation, we take the
top-pion mass $m_{\pi_{t}}$ and the mixing factor $
k=k_{\tau\mu}=k_{\tau e}$ as free parameters.

We plot the branching ratios $Br(\tau\rightarrow e\gamma)$ and
$Br(\mu\rightarrow e\gamma)$ as function of the mixing factor $k$
for three values of the top-pion mass in Fig.2 and Fig.3,
respectively. To compare the value of $Br(\mu\rightarrow e\gamma)$
given by $\pi_{t}^{0}$ exchange with its current experimental
limit, we have used the horizontal solid line to denote
$Br(\mu\rightarrow e\gamma)= 1.1\times 10^{-11}$ in Fig.3. One can
see from Fig.2 and Fig.3 that the branching ratios of the $LFV$
processes $l_{i}\rightarrow l_{j}\gamma$ are $Br(\tau\rightarrow
\mu\gamma)\simeq Br(\tau\rightarrow e\gamma)<2\times10^{-10}$ and
$Br(\mu \rightarrow e\gamma)<1.4\times10^{-7}$ in most of the
parameter space of $TC2$ models. The branching ratio
$Br(\tau\rightarrow l\gamma)$ is at least four orders of magnitude
below the present experimental bound on $\tau\rightarrow l\gamma$,
which is far from the reach of present or next generation
experiments. The branching ratio $Br(\mu \rightarrow e\gamma)$ can
be above or below the present experimental bound on
$\mu\rightarrow e\gamma$, which depends on the value of the
top-pion mass $m_{\pi_{t}}$ and the mixing factor $k$. In most of
the parameter space, the value of the $Br(\mu\rightarrow e\gamma)$
is in the range of $2.5\times 10^{-12}\sim 1.4\times 10^{-7}$.

\begin{figure}[htb]
\vspace{-1cm}
\begin{center}
\epsfig{file=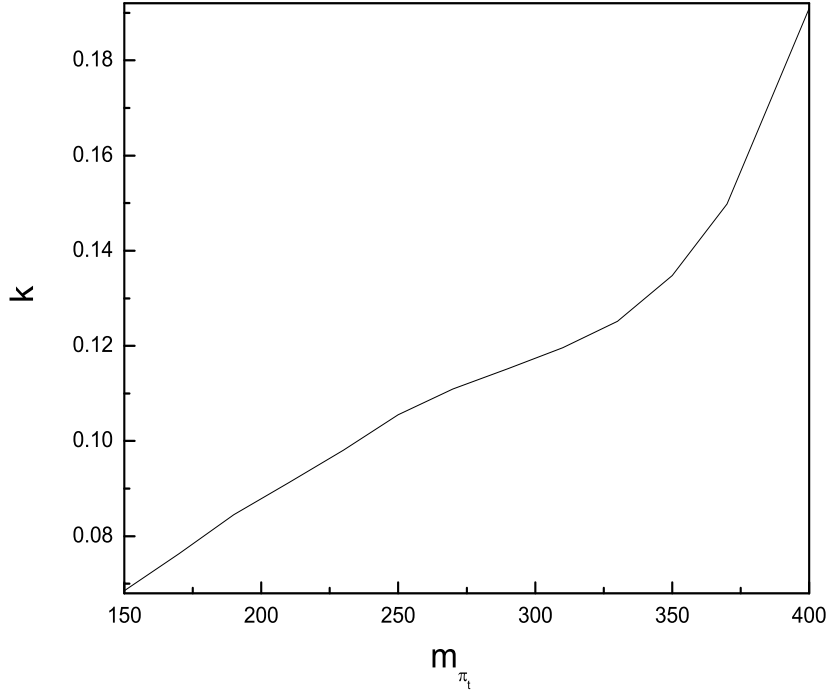,width=350pt,height=300pt}\vspace{-1.0cm}
\hspace{5mm} \caption{The flavor mixing factors $k$ as a function
of top-pion mass $m_{\pi_{t}}$ for
\hspace*{2.2cm}$Br({\mu\rightarrow e\gamma})=1.2\times 10^{-11}$.}
\label{ee}
\end{center}
\end{figure}

Using the present experimental bound on the LFV process
$\mu\rightarrow e\gamma$, we can give the constraints on the
mixing factor $k$ for $150GeV \leq m_{\pi_{t}}\leq 400GeV$. The
numerical results are showed in Fig.4. From Fig.4 we can see that
the mixing factor $k$ increases as $m_{\pi_{t}}$ increasing. If we
demand that the top-pion mass is smaller than $400GeV$, then there
must be $k\leq 0.21$. Thus, the present experimental upper bound
of the $LFV$ process $\mu\rightarrow e\gamma$ gives severe
constraints on the free parameters $m_{\pi_{t}}$ and $k$ of $TC2$
models. In the next section, we will take the $\mu\rightarrow
e\gamma$ constraints into account and calculate the branching
ratios of the $LFV$ processes $l_{i}\rightarrow l_{j}l_{k}l_{l}$.

\vspace{0.5cm}
\noindent{\bf3. The $LFV$ processes
$l_{i}\rightarrow l_{j}l_{k}l_{l}$}

In $TC2$ models, the $LFV$ processes $l_{i}\rightarrow
l_{j}l_{k}l_{l}$ can be generated at the tree level and also can
be induced via photon penguin diagrams at the one-loop level, as
shown in Fig.5. For the diagrams Fig.5 (b),(c),(d), we have taken
$k=l$.

\begin{figure}[htb]
\vspace{-8cm}
\begin{center}
\epsfig{file=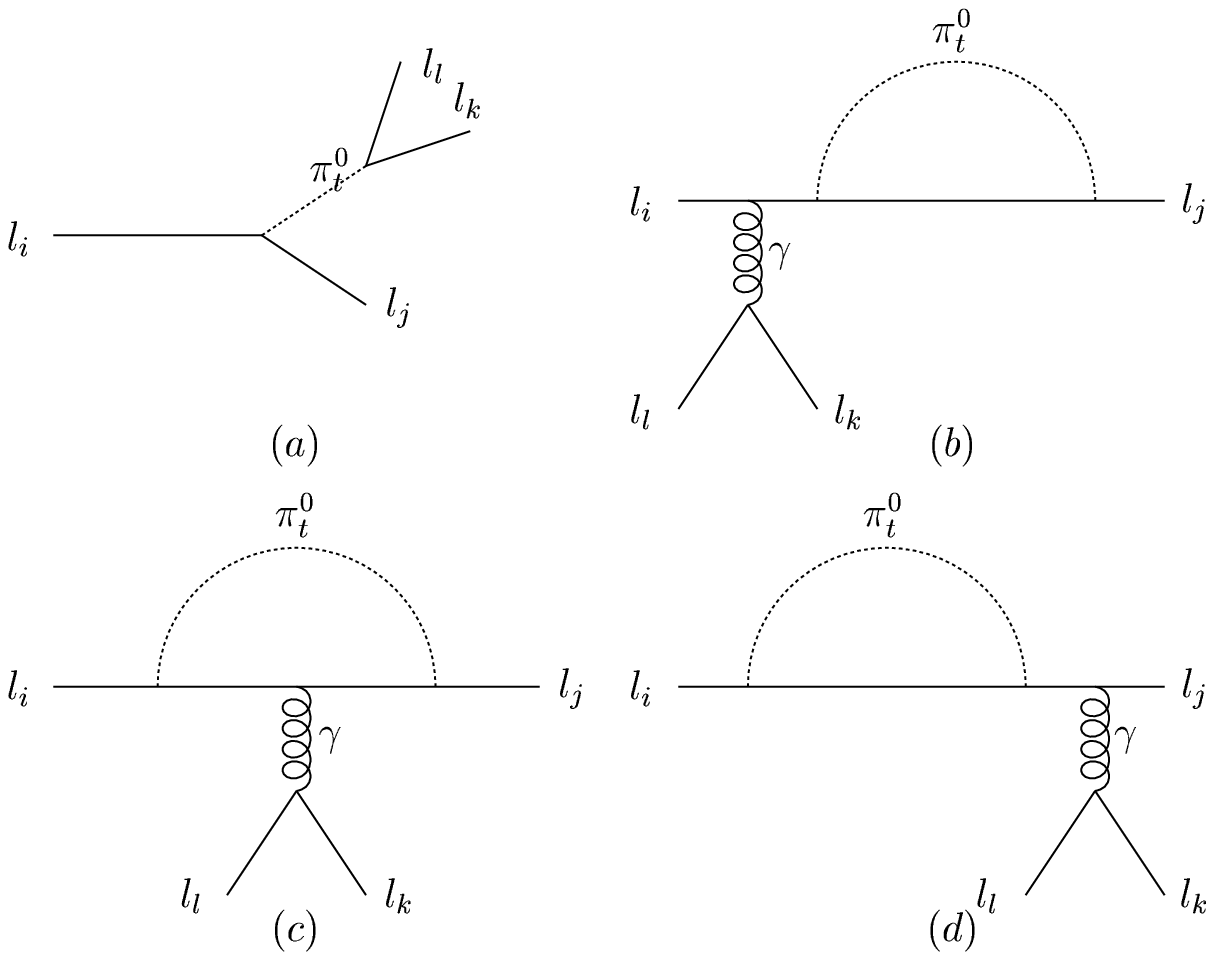,width=550pt,height=700pt} \vspace{-8.0cm}
\hspace{5mm} \caption{The tree-level and one-loop Feynman diagrams
contribute to the $LFV$ processes \hspace*{1.8cm}$l_{i}\rightarrow
l_{j}l_{k}l_{l}$ induced by $\pi^{0}_{t}$ exchange.} \label{ee}
\end{center}
\end{figure}

Let us first consider the contributions of the neutral top-pion
$\pi^{0}_{t}$ to the $LFV$ processes $l_{i}\rightarrow
l_{j}l_{k}l_{l}$ via Fig.5(a). For the decay $\mu\rightarrow 3e$,
it is induced by the $FC$ scalar coupling
$\pi^{0}_{t}\overline{\mu}e$. However, the topcolor interactions
only contact with the third generation fermions. The flavor mixing
between the first and second generation fermions is very small
[15]. In numerical estimation, we will assume $k_{\mu e}\simeq 0$.
So, the branching ratio $Br(\mu\rightarrow 3e)$ induced by
$\pi^{0}_{t}$ exchange at the tree-level is zero. The $LFV$
processes $\tau\rightarrow 2\mu e$ and $\tau\rightarrow 2e\mu$ can
only be generated via the $FC$ couplings $\pi^{0}_{t}\tau e$ and
$\pi^{0}_{t}\tau\mu$. The decay widths of the processes
$\tau\rightarrow l_{i}l_{j}l_{k}$ are given by:
\begin{eqnarray}
\Gamma(\tau\rightarrow
3e)&=&\frac{m^{7}_{\tau}m^{2}_{e}}{1042\pi^{3}m_{\pi}^{4}\nu^{4}}k^{2},\\
\Gamma(\tau\rightarrow
3\mu)&=&\frac{m^{7}_{\tau}m^{2}_{\mu}}{1042\pi^{3}\nu^{4}m^{4}_{\pi}}k^{2},\\
\Gamma(\tau\rightarrow
2\mu e)&=&\frac{m^{7}_{\tau}m^{2}_{\mu}}{3072\pi^{3}m_{\pi}^{4}\nu^{4}}k^{2},\\
\Gamma(\tau\rightarrow
2e\mu)&=&\frac{m^{7}_{\tau}m^{2}_{e}}{3072\pi^{3}m_{\pi}^{4}\nu^{4}}k^{2},
\end{eqnarray}
where $m_{l}$($l=\mu$,$e$, or $\tau$) represents the lepton mass
and $k=k_{\tau\mu}=k_{\tau e}$.

Now we consider the one-loop contributions of the neutral top-pion
$\pi^{0}_{t}$ to the $LFV$ processes $l_{i}\rightarrow
l_{j}l_{k}l_{l}$ via the photonic penguin diagrams shown in
Fig.5(b)(c)(d). Same as Fig.1, the internal fermion line of the
photonic penguin diagrams only is the $\tau$ fermion line.
Comparing Fig.5 with Fig.1, one can use the branching ratios
$Br(l_{i}\rightarrow l_{j}\gamma)$ to express the branching ratios
$Br(l_{i}\rightarrow l_{j}l_{k}l_{l})$[10]. The one-loop
expressions of the branching ratios $Br(l_{i}\rightarrow
l_{j}l_{k}l_{l})$ can be written as:
\begin{eqnarray}
Br^{1-loop}(\tau\rightarrow 3\mu)&=&Br^{1-loop}(\tau\rightarrow
3e)=Br^{1-loop}(\tau\rightarrow 2\mu
e)=Br^{1-loop}(\tau\rightarrow 2e\mu)\nonumber\\
&\simeq&\frac{\alpha_{e}}{3\pi}(ln\frac{m^{2}_{\tau}}{m^{2}_{\mu}}-\frac{11}{4})
Br(\tau\rightarrow
e\gamma),\\
Br^{1-loop}(\mu\rightarrow
3e)&\simeq&\frac{\alpha_{e}}{3\pi}(ln\frac{m^{2}_{\tau}}{m^{2}_{\mu}}-\frac{11}{4})
Br(\mu\rightarrow
e\gamma).
\end{eqnarray}

For the $LFV$ processes $\tau\rightarrow l_{i}l_{j}l_{k}$, we have
assumed $m_{\mu}\simeq0$, $m_{e}\simeq0$ and taken
$Br(\tau\rightarrow e\gamma)\simeq Br(\tau\rightarrow\mu\gamma)$.
The expressions of the $Br(\tau\rightarrow e\gamma)$ and
$Br(\mu\rightarrow e\gamma)$ have been given in Eq.(16) and
Eq.(17), respectively.

Comparing the one-loop contributions of $\pi^{0}_{t}$ to the
$l_{i}\rightarrow l_{j}l_{k}l_{l}$ with the tree-level
contributions, we find that the branching ratios
$Br(\tau\rightarrow 2e\mu)$, $Br(\tau\rightarrow 3e)$, and
$Br(\mu\rightarrow 3e)$ given by the one-loop diagrams mediated by
 $\pi^{0}_{t}$ exchange are larger than those generated by the
tree-level diagrams at least by four orders of magnitude. This is
because the FD coupling $\pi^{0}_{t}ee$ is proportional to
$\frac{m_{e}}{\nu}$, which can strongly suppress the values of
these branching ratios. However, for the processes
$\tau\rightarrow3\mu$ and $\tau\rightarrow2\mu e$, two kinds of
contributions are comparable. Thus, we will ignore the tree-level
contributions of $\pi^{0}_{t}$ exchange to the $\tau\rightarrow
3e$, $\tau\rightarrow 2e\mu$ and $\mu\rightarrow 3e$ in the
following numerical estimation.

\begin{figure}[htb]
\vspace{-0.5cm}
\begin{center}
\epsfig{file=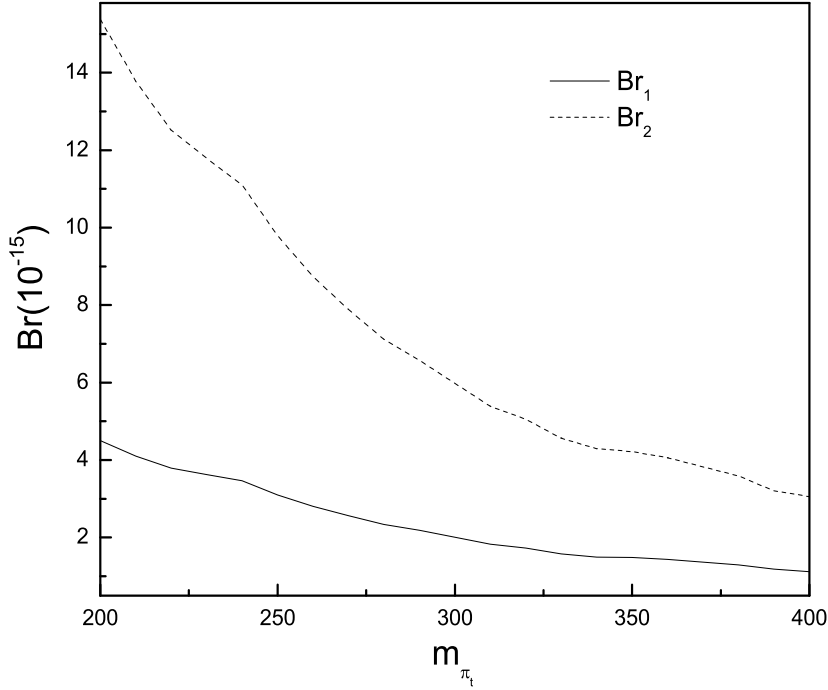,width=350pt,height=300pt} \vspace{-1.0cm}
\hspace{5mm} \caption{The branching rations $ Br(l_{i}\rightarrow
l_{j}l_{k}l_{l})$ as functions of the top-pion mass $m_{\pi_{t}}$.
\hspace*{2.0cm} We have assumed $Br_{1}=Br(\tau\rightarrow 3e)$
and $Br_{2}=Br(\tau \rightarrow 3\mu)$.} \label{ee}
\end{center}
\end{figure}

Taking into account the constraints of the current experimental
upper bound $Br(\mu\rightarrow e\gamma)\leq1.1\times10^{-11}$ on
the free parameters $m_{ \pi_{t}}$ and $k$, we find that the
branching ratio $Br(\mu\rightarrow 3e)$ is approximately equal to
$2.87\times10^{-14}$, which might be observable in the planned
experiments of the next generation. Combining the tree-level and
one-loop contributions, we have $Br(\tau\rightarrow 3e)\simeq
Br(\tau\rightarrow 2e\mu)$ and $Br(\tau\rightarrow 3\mu)\simeq
Br(\tau\rightarrow 2\mu e)$ in $TC2$ models, which are shown in
Fig.6 as functions of the top-pion mass $m_{ \pi_{t}}$. In Fig.6,
we have used $Br_{1}$ and $Br_{2}$ represent $Br(\tau\rightarrow
3e)$ and $Br(\tau\rightarrow 3\mu)$, respectively. One can see
from Fig.6 that the branching ratios slowly decrease as $m_{
\pi_{t}}$ increasing. As long as $m_{ \pi_{t}}<400GeV$, we have
$Br(\tau\rightarrow 3e)\simeq Br(\tau\rightarrow 2e\mu)\geq
1.1\times 10^{-15}$ and $Br(\tau\rightarrow 3\mu)\simeq
Br(\tau\rightarrow 2\mu e)\geq 3.1\times 10^{-15}$. Even we take
$m_{ \pi_{t}}=200GeV$, the branching ratio $Br(\tau\rightarrow
3\mu)$ can only reach $1.54\times10^{-14}$, which is far below the
experimental bound on $\tau\rightarrow l_{i}l_{j}l_{k}$ ($10^{-6}$
or $10^{-7}$) [18,21].

\vspace{.5cm} \noindent{\bf4. Conclusions}

The presence of physics top-pions in the low-energy spectrum is a
common feature of topcolor models. The physics top-pions have
large Yukama couplings to the third family fermions and can induce
the $FC$ scalar couplings, which might give significant
contributions to the $FC$ processes. The effects of top-pion on
these processes are governed by its mass $m_{\pi_{t}}$ and the
relevant flavor mixing factors.

In this paper we study the contributions of the neutral top-pion
$\pi^{0}_{t}$ predicted by $TC2$ models on the $LFV$ processes
$l_{i}\rightarrow l_{j}\gamma$ and $l_{i}\rightarrow
l_{j}l_{k}l_{l}$. We find that the branching ratio
$Br(\tau\rightarrow e\gamma)$ is approximately equal to the
branching ratio $Br(\tau\rightarrow \mu\gamma)$ and is smaller
than $2\times10^{-10}$ in all parameter space of $TC2$ models. The
present experimental bound on $\mu\rightarrow e\gamma$ produces
severe constraints on the top-pion mass $m_{ \pi_{t}}$ and the
mixing factor $ k $. Based on these constraints on the free
parameters of $TC2$ models, we further calculate the contributions
of $\pi^{0}_{t}$ to the $LFV$ processes $l_{i}\rightarrow
l_{j}l_{k}l_{l}$ at the tree-level and one-loop level. For the
$LFV$ processes $\mu\rightarrow 3e$, $\tau\rightarrow 3e$, and
$\tau\rightarrow 2e\mu$, the contributions coming from photonic
penguin diagrams are larger than those from the tree-level
top-pion exchange at least by four orders of magnitude. While two
kinds of contributions are comparable for the processes
$\tau\rightarrow 3\mu$ and $\tau\rightarrow 2\mu e$. If we take
the top-pion mass $m_{ \pi_{t}}\leq 400GeV$, we have
$Br(\tau\rightarrow 3e)\simeq Br(\tau\rightarrow 2e\mu)\geq
1.1\times 10^{-15}$ and $Br(\tau\rightarrow 2\mu e)\simeq
Br(\tau\rightarrow 3\mu)\geq 3.1\times 10^{-15}$, which can not be
observable in the near future experiments. However, the branching
ratio $Br(\mu\rightarrow 3e)$ is approximately equal to $2.8\times
10^{-14}$, which may be observable in the planned experiments of
the next generation.

\vspace{1.5cm} \noindent{\bf Acknowledgments}

This work was supported  by the National Natural Science
Foundation of China (90203005).


\vspace{2.5cm}
\begin{thebibliography}{99}
\bibitem{y1}C. T. Hill, {\em Phys. Lett. B}{\bf 345}, 483(1995);
            K. Lane and E. Eichten, {\em Phys. Lett. B}{\bf 352}, 383(1995);
            K. Lane, {\em Phys. Lett. B}{\bf 433}, 96(1998);
            G. Cvetic, {\em Rev. Mod. Phys.} {\bf 71}, 513(1999).
\bibitem{y2}M. B. Popovic, E. H. Simmons, {\em Phys. Rev. D}{\bf
             58}, 095007(1998); G. Burdman and N. Evans, {\em Phys. Rev. D}{\bf
             59}, 115005(1999).
\bibitem{y3}B. A. Dobrescu, C. T. Hill, {\em Phys. Rev.
             Lett.} {\bf 81}, 2634(1998); R. S. Chivukula, B. A. Dobrescu, H.
             Georgi, C.T. Hill, {\em Phys. Rev. D}{\bf 59}, 075003(1999).
\bibitem{y4}H.-J. He, T. M. P. Tait, C.-P. Yuan, {\em Phys. Rev. D}{\bf
             62}, 011702(2000).
\bibitem{y5}Super-Kamiokande Collaboration, S. Fukuda et al.,
            {\em Phys. Rev. Lett.} {\bf 86}, 5651(2001);
            5656(2001); SNO Collaboration, Q. R. Ahmad et al.,
            {\em Phys. Rev. Lett.} {\bf 87}, 071301(2001); {\bf 89}, 011301(2002);
            {\bf 89}, 011302(2002); Homestake Collaboration,
            R. Davis, {\em Rev. Mod. Phys.} {\bf 75}, 985(2003).
\bibitem{y6}Super-Kamiokande Collaboration, Y. Fukuda et al.,
            {\em Phys. Rev. Lett.} {\bf 81}, 1562(1998); {\bf
            82}, 2644(1999); {\bf 85}, 3999(2000).
\bibitem{y7}KamLAND Collaboration, K. Eguchi et al., {\em\ Phys. Rev.
            Lett.} {\bf 90}, 021802(2003); K2K Collaboration, M.
            H. Ahn et al., {\em Phys. Rev. Lett.} {\bf 90}, 041801(2003).
\bibitem{y8}S. Nussinov, R. D. Peccei, X. M. Zhang, {\em Phys. Rev.
            D}{\bf 63}, 016003(2000); D. Delepine, F. Vissani, {\em Phys. Lett.
            B}{\bf 522}, 95(2001); D. Black, T. Han, H.-J. He, M. Sher,
            {\em Phys. Rev. D}{\bf 66}, 053002(2002); E. Ma, {\em Nucl. Phys. Proc.
            Suppl.} {\bf 123}, 125(2003).
\bibitem{y9}A. Ilakovac, A. Pilaftsis, {\em Nucl. Phys. B}{\bf
            437}, 491(1995); A. Ilakovac, {\em Phys. Rev. D}{\bf
            62}, 036010(2000); D. A. Dicus, H. -J. He, J. N. Ng, {\em Phys. Rev.
            Lett.} {\bf 87}, 111803(2001); J. I. Illana, J. Riemann,
            {\em Phys. Rev. D}{\bf 63}, 053004(2001); G. Cvetic, C.
            Dib, C. S. Kim, J. D. Kim, {\em Phys. Rev. D}{\bf
            66}, 034008(2002);
\bibitem{y10}J. L. Feng, Y. Nir, Y. Shadmi, {\em Phys . Rev. D}{\bf
             61}, 113005(2000); Y. Okada, K. Okumura, Y. Shimizu,
             {\em Phys. Rev. D}{\bf 61}, 094001(2000); K. S. Babu,
             C. Kolda, {\em Phys. Rev. Lett.} {\bf
             89}, 241802(2002); M. Sher, {\em Phys. Rev. D}{\bf
             66}, 05731(2002); J. Hisano, hep-ph/0209005; R. A.
             Diaz, R. Martinez, N. Poreda, hep-ph/0209122; A.
             Dedes, J. R. Ellis, M. Raidal, {\em Phys. Lett. B}{\bf
             549}, 159(2002); A. Gemintern, S. Bar-Shalom and G.
             Eilam, {\em Phys. Rev. D}{\bf 67}, 115012(2003); B. A.
             Campbell, D. W. Maybury and B. Murakami,
             hep-ph/0311244;
\bibitem{y11}M. Sher and Y. Yuan, {\em Phys. Rev. D}{\bf
             44}, 1461(1991); S. K. Kang and K. Y. Lee, {\em Phys. Lett.
             B}{\bf 521}, 61(2001); E. O. Iltan, {\em Phys. Rev.
             D}{\bf 64}, 115005(2001); E. O. Iltan and I. Turan, {\em Phys. Rev.
             D}{\bf 65}, 013001(2002); Rodolfo A. Diaz, R.
             Martinez, and J. Alexis Rodriguez, {\em Phys. Rev.
             D}{\bf 67}, 075011(2003); Rodolfo A. Diaz, R.
             Martinez, and Carlos E. Sandoval, hep-ph/0311201.
\bibitem{y12}A. Ghosal, Y. Koide, H. Fusaoka, {\em Phys. Rev. D}{\bf 64}, 053012(2001);
             E. Mituda, K. Sasaki, hep-ph/0103202.
\bibitem{y13}T. Rador, {\em Phys. Rev. D}{\bf 60}, 095012(1999);
             Chongxing Yue, Guoli Liu, Jiantao Li, {\em Phys. Lett.
             B}{\bf 496}, 89(2002); Chongxing Yue, Yanming Zhang,
             Lanjun Liu, {\em Phys. Lett. B}{547}, 252(2002);
             Chongxing Yue, Lanjun Liu, {\em Phys. Lett. B}{\bf
             564}, 55(2003).
\bibitem{y14}G. Burdman and D. Kominis, {\em Phys. Lett. B}{\bf
             403}, 101(1997); W. Loinaz and T. Takeuchi, {\em Phys. Rev.
             D}{\bf 53}, 015005(1999).
\bibitem{y15}G. Burdman, {\em Phys. Rev. Lett.} {\bf 83}, 2888(1999);
             H.-J. He, C.-P. Yuan, {\em Phys. Rev. Lett.} {\bf
             83}, 28(1999); H.-J. He, S. Kanemura, C.-P. Yuan,
             {\em Phys. Rev. Lett.} {\bf 89}, 101803(2002).
\bibitem{y16}Chongxing Yue, Lanjun Liu, Dongqi Yu, {\em Phys. Rev.
             D}{68}, 035002(2003).
\bibitem{y17}S. M. Bilenky, S. T. Petcov and B. Pontecorvo, {\em Phys. Lett.
             B}{\bf 67}, 309(1977); S. T. Pectcov, {\em Sov. J. Nucl.
             Phys.} {\bf 25}, 340(1997); T. P. Cheng and L. -F. Li, {\em Phys. Rev.
             Lett.} {\bf 45}, 1908(1980).
\bibitem{y18}D. E. Groom et al.[Particle Data Group], {\em Eur. Phys. J.
             C}{\bf 15}, 1(2000); K. Hagiwara et al., {\em Phys. Rev. D}{\bf
             66}, 010001(2002) and references therein.
\bibitem{y19}L. M. Barkov et al., {\em Research Proposal for
             experiment at PSI(1999), http://meg.psi.ch}.
\bibitem{y20}C. T. Hill and E. H. Simmons, {\em Phys. Rept.} {\bf
             381}, 235(2003), [Erratum -ibid, {\bf 390},
             553(2004)].
\bibitem{y21}T. Hokuue, {\em Talk in PDF 2002, 24-28 May 2002,
             Williamsbury, Virginia, USA; Y. Yusa, Talk in PDF
             2002, 24-28 May 2002, Williamsbury, Virginia, USA.}
\end{thebibliography}
\end{document}